\documentclass[aps,prl,twocolumn, 10pt]{revtex4-1}

\usepackage{graphics,graphicx, array, afterpage}
\usepackage{qcircuit,amsmath}
\usepackage{grffile}
\usepackage[export]{adjustbox}
\usepackage{hyperref}
\usepackage{lineno}
\usepackage{sidecap}
\usepackage[small,compact]{titlesec} 
\usepackage{braket}
\usepackage{multirow}
\newcommand{\Gate}[1]{\textsc{#1}}
\newcommand{\rzgate}{{\Gate{R}}_{z}}

\newcommand{\rygate}{{\Gate{R}}_{y}}
\newcommand{\cnotgate}{\Gate{cnot}} 
\newcommand{\swapgate}{\Gate{swap}} 
\newcommand{\xxgate}{\Gate{XX}}

\begin{document}

\title{Quantum Computer Systems for Scientific Discovery}
\author{Yuri Alexeev$^1$}
\author{Dave Bacon$^2$}
\author{Kenneth R. Brown$^{3,4,5}$}
\author{Robert Calderbank$^{3,6}$}
\author{Lincoln D. Carr$^7$}
\author{Frederic T. Chong$^8$}
\author{Brian DeMarco$^9$}
\author{Dirk Englund$^{10}$}
\author{Edward Farhi$^{11,12}$}
\author{Bill Fefferman$^8$}
\author{Alexey V.\ Gorshkov$^{13,14}$}
\author{Andrew Houck$^{15}$}
\author{Jungsang Kim$^{3,5,16}$}
\author{Shelby Kimmel$^{17}$}
\author{Michael Lange$^{18}$}
\author{Seth Lloyd$^{19}$}
\author{Mikhail D. Lukin$^{20}$}
\author{Dmitri Maslov$^{21}$}
\author{Peter Maunz$^{22}$}
\author{Christopher Monroe$^{*\ 13,16}$}
\author{John Preskill$^{23}$}
\author{Martin Roetteler$^{24}$}
\author{Martin J. Savage$^{25}$}
\author{Jeff Thompson$^{15}$}

\affiliation{$^{1}$Argonne National Laboratory, Lemont, IL}
\affiliation{$^{2}$Google, Inc., Seattle, WA}
\affiliation{$^{3}$Department of Electrical and Computer Engineering, Duke University, Durham, NC}
\affiliation{$^{4}$Department of Chemistry, Duke University, Durham, NC}
\affiliation{$^{5}$Department of Physics, Duke University, Durham, NC}
\affiliation{$^{6}$Department of Computer Science and Department of Mathematics, Duke University, Durham, NC}
\affiliation{$^{7}$Department of Physics, Colorado School of Mines, Golden, CO}
\affiliation{$^{8}$Department of Computer Science, University of Chicago, Chicago, IL}
\affiliation{$^{9}$Department of Physics and IQUIST, University of Illinois, Urbana-Champaign, IL}
\affiliation{$^{10}$Department of Electrical Engineering and Computer Science, Massachusetts Institute of Technology, Cambridge, MA}
\affiliation{$^{11}$Google, Inc., Venice CA}
\affiliation{$^{12}$Department of Physics, Massachusetts Institute of Technology, Cambridge, MA}
\affiliation{$^{13}$Joint  Quantum  Institute, Joint Center for Quantum Information and Computer Science, and Department of Physics, University  of  Maryland,  College  Park,  MD}
\affiliation{$^{14}$National Institute of Standards and Technology, Gaithersburg, MD}
\affiliation{$^{15}$Department of Electrical Engineering, Princeton University, Princeton, NJ}
\affiliation{$^{16}$IonQ, Inc., College Park, MD}
\affiliation{$^{17}$Department of Computer Science, Middlebury College, Middlebury, VT}
\affiliation{$^{18}$L3Harris Technologies, Melborune, FL}
\affiliation{$^{19}$Department of Mechanical Engineering, Massachusetts Institute of Technology, Cambridge, MA}
\affiliation{$^{20}$Department of Physics, Harvard University, Cambridge, MA}
\affiliation{$^{21}$IBM T.J. Watson Research Center, Yorktown Heights, NY}
\affiliation{$^{22}$Sandia National Laboratories, Albuquerque, NM}
\affiliation{$^{23}$Institute for Quantum Information and Matter and Walter Burke Institute for Theoretical Physics, California Institute of Technology, Pasadena CA}
\affiliation{$^{24}$Microsoft Quantum, Redmond, WA}
\affiliation{$^{25}$Institute for Nuclear Theory and Department of Physics, University of Washington, Seattle, WA}

\begin{abstract} 
\vspace{\baselineskip}

\noindent
The great promise of quantum computers comes with the dual challenges of building them and finding their useful applications.  We argue that these two challenges should be considered together, by co-designing full-stack quantum computer systems along with their applications in order to hasten their development and potential for scientific discovery. In this context, we identify scientific and community needs, opportunities, a sampling of a few use case studies, and significant challenges for the development of quantum computers for science over the next 2--10 years.

\vspace{\baselineskip}

\noindent
\textit{This document is written by a community of university, national laboratory, and industrial researchers in the field of Quantum Information Science and Technology, and is based on a summary from a U.S. National Science Foundation workshop on Quantum Computing held on October 21--22, 2019 in Alexandria, VA.} 
 
\vspace{\baselineskip}
\noindent
* to whom correspondence should be addressed: monroe@umd.edu

\vspace{5cm}  
\end{abstract}

\maketitle


\pagebreak[4]

\section{Executive Summary}

The quantum computer promises enormous information storage and processing capacity that can eclipse any possible conventional computer \cite{MikeAndIke}.  
This stems from quantum entangled superpositions of individual quantum systems, usually expressed as a collection of quantum bits or qubits, whose full description requires exponentially many parameters. 
However, there are two critical challenges in bringing quantum computers to fruition:

\textbf{Challenge 1:} The vast amount of information contained in massively entangled quantum systems is itself not directly accessible, due to the reduction of quantum superpositions upon measurement. Instead, useful quantum computer algorithms guide a quantum state to a much simpler form to produce some type of global property of the information that can be directly measured. However, the full scope of applications that can exploit entangled superpositions in this way and how exactly quantum computers will be used in the future remains unclear.
    
\textbf{Challenge 2:} Quantum computers are notoriously difficult to build, requiring extreme isolation of a large number of individual qubits, while also allowing exquisite control of their quantum states and high-accuracy measurements. Quantum computer technology is nothing like classical computer hardware and involves unconventional information carriers in exotic environments like high vacuum or very low temperature.  Ultimately, large-scale quantum computers will utilize error-correction techniques that are much more complex than their classical counterparts.

This article combines these two challenges by promoting the idea of \textit{co-designing} quantum computers with their scientific applications in a vertically-integrated approach that addresses scientific opportunities at all levels of the quantum computer stack.  

Since the birth of quantum computing in the 1990s, there has been enormous progress in the isolation and control of good qubit platforms \cite{Ladd:2010}. 
Some of these quantum technologies, based on individual atoms controlled by laser beams \cite{Monroe:2013, Saffman:2019} or superconducting circuitry coupled with microwave fields \cite{Devoret:2013}, are now being built into small systems. 
This has led to a new era of quantum computing, paralleling the transition from transistors to integrated circuits many decades ago, which is expected to lead to significant scientific opportunities. 
In this position paper, we, therefore, do not focus on the physics or development of qubit technologies at the component level. We also do not speculate on new qubit technologies that may emerge in the coming years. These are important and foundational research activities, but they are also typically divorced from systems-level considerations of operating quantum computers. 

Here we concentrate on the near-term prospects and scientific opportunities generated by an integrated consideration of the complete quantum computer stack using existing quantum system technologies.  
At the bottom of the stack, device-specific qubit control considerations will impact the engineering of native interactions or gate sets, connectivity, and thus net performance on high-level quantum computer applications.
In the middle of the stack, compilation of native gates into standard gate sets and higher-level quantum subroutines can be compressed and compiled further using techniques from quantum computer science.  Quantum error-correction encoding of qubits, a relatively new field in its own right, will become ever more important as the systems grow in the number of qubits and their circuit depth. At the top of the stack, the execution of quantum circuits can simulate difficult quantum problems, from molecular properties, chemistry, and materials science \cite{McClean:2016} to nuclear and particle physics models beyond the reach of conventional computers \cite{Preskill_simulating_2018}.  They may also find use as general optimizers of generic models \cite{Farhi:2014} that could be applied to logistics, economics, and climate science.  

We advocate for the continual development and operation of multiple generations of quantum computer \textbf{\textit{systems}} using current technology specifically designed with the above types of opportunities in mind.  We propose iterating between building and using the devices, with a full-stack scientific mission that can be fulfilled by a unified effort at universities, at national laboratories, and in industry. As future qubit technologies are developed, we expect they will be integrated into the stack.
While there is growing industrial interest in building quantum computers, we note that these efforts may not be focused on co-designing quantum computers that will have sufficient flexibility to address scientific opportunities at the various levels of the quantum computer stack.
Over the next 2--10 years, the research and development approach we propose here is expected to generate new science, stimulate the transition of academic and national laboratory programs in quantum computing to industry, and also train future quantum engineers.
The overriding high-level aim of this proposed path is to hasten the development of a wide range of concrete scientific applications for quantum computers, with parallel efforts in quantum simulation \cite{Carr:2019} and quantum communication \cite{Loncar:2019}.

\section{Introduction}     

Quantum computers represent a fundamental departure from the way we process information. At its core, a quantum computer consists of quantum bits (qubits) or equivalent quantum information carriers, that allow the storage and processing of quantum superpositions of data.  
A single qubit $\ket{x_i}$ is a quantum two-level system that can store a superposition of both $x_i=0$ and $x_i=1$. A collection of $n$ qubits can be represented by a quantum state $\ket{\Psi}$ that stores an arbitrary \textit{entangled} superposition of all $n$-bit binary numbers,
\begin{eqnarray} \label{eqn:state}
\ket{\Psi} &=& \sum_{k=0}^{2^n-1}\alpha_k \ket{k} \\
&=&\sum_{x_i \in \{0,1\}}\alpha_{x_{n-1}x_{n-2}\ldots x_0} \ket{x_{n-1}}\ket{x_{n-2}}\ldots\ket{x_0},
\end{eqnarray}
where the $2^n$ weights or amplitudes of each basis state $\ket{k}$ are given by complex numbers $\alpha_k$,
whose index $k$ can also be expressed as a string of bit values $x_{n-1}x_{n-2}...x_0$.
The superposition amplitudes $\alpha_k$ evolve according to the unitary (time-reversible) Schr\"odinger wave equation that dictates how the amplitudes of the various basis states can be controlled, governed by an underlying Hamiltonian or energy functional.
When the qubits are measured, they assume definite values with probabilities $P(k) = |\alpha_k|^2$ given by the corresponding amplitudes of the underlying quantum state.
The entanglement of the above general quantum superposition represents a complex web of natural links between qubits, giving rise to a network of correlations while maintaining the character of superposition within individual qubits.  
Quantum entanglement allows an efficient ``wiring" of qubit states without any real wires or physical connections between the qubits, and it has no classical analog.

There are several known quantum algorithms that offer various advantages or speedups over classical computing approaches, some even reducing the complexity class of the problem.  
These algorithms generally proceed by controlling the quantum interference between the components of the underlying entangled superpositions in such a way that only one or relatively few quantum states have a significant amplitude in the end.  A subsequent measurement can, therefore, provide global information on a massive superposition state with significant probability.

The coherent manipulation of quantum states that defines a quantum algorithm can be expressed through different quantum computational modes with varying degrees of tunability and control.  The most expressive quantum computing mode presently known is the universal gate model, similar to universal gate models of classical computation. Here, a quantum algorithm is broken down to a sequence of modular quantum operations or gates between individual qubits. There are many universal quantum gate families operating on single and pairwise qubits \cite{DiVincenzo:1995}, akin to the NAND gate family in classical computing. 
One popular universal quantum gate family is the grouping of two-qubit $\cnotgate$ gates on every pair of qubits along with rotation gates on every single qubit \cite{MikeAndIke}, as displayed in Fig. \ref{fig:gates}. 
With universal gates, an arbitrary entangled state and thus any quantum algorithm can be expressed. Alternative modes such as measurement-based or cluster-state quantum computing \cite{Raussendorf:2003} can be shown to be formally equivalent to the universal gate model.
Another interesting quantum model worth mentioning is “quantum walks" model \cite{solenov2019quantum}, which is a natural mathematical framework to construct quantum gates in qubit systems controlled via pulses.
Like the NAND gate in classical CMOS technology, the particular choice of universal gate set or even mode of quantum computing is best determined by the quantum hardware itself and its native interactions and available controls.  The structure of the algorithm itself may also impact the optimal choice of gate set or quantum computing mode. 

\begin{figure}[h] 
  \begin{minipage}[b]{0.49\linewidth}
  \begin{eqnarray*} \\ \\
    \ket{0} &\longrightarrow& \cos\frac{\theta}{2}\ket{0}-ie^{+i\phi}\sin\frac{\theta}{2}\ket{1} \\ 
    \ket{1} &\longrightarrow& \cos\frac{\theta}{2}\ket{1}-ie^{-i\phi}\sin\frac{\theta}{2}\ket{0} \\
  \end{eqnarray*}
   \vspace{\baselineskip}
  \mbox{\small \Qcircuit @C=2em @R=1.4em {\lstick{\ket{x}}& \gate{R(\theta,\phi)} & \rstick{\ket{\tilde{x}}}\qw}} \\
  \vspace{0.5\baselineskip}   (a)
  \end{minipage}
  \begin{minipage}[b]{0.49\linewidth}
    \begin{eqnarray*}
        \ket{0}\ket{0} &\longrightarrow& \ket{0}\ket{0} \\ 
        \ket{0}\ket{1} &\longrightarrow& \ket{0}\ket{1} \\ 
        \ket{1}\ket{0} &\longrightarrow& \ket{1}\ket{1} \\ 
        \ket{1}\ket{1} &\longrightarrow& \ket{1}\ket{0} 
    \end{eqnarray*}
   \mbox{\small \Qcircuit @C=2em @R=1.4em {\lstick{\ket{x_C}}& \ctrl{1} & \rstick{\ket{\tilde{x}_C}}\qw \\ \lstick{\ket{x_T}}& \targ & \rstick{\ket{\tilde{x}_T}}\qw}} \\
  \vspace{0.5\baselineskip}   (b)
  \end{minipage}
\caption{\label{fig:gates}The rotation and controlled-NOT ($\cnotgate$) gates are an example of a universal quantum gate family when available on all qubits, with explicit evolution (above) and quantum circuit block schematics (below). (a) The single-qubit rotation gate $R(\theta,\phi)$, with two continuous parameters $\theta$ and $\phi$, evolves input qubit state $\ket{x}$ to output state $\ket{\tilde{x}}$.  (b) The $\cnotgate$ (or reversible XOR) gate on two qubits evolves two (control and target) input qubit states $\ket{x_C}$ and $\ket{x_T}$ to output states $\ket{\tilde{x}_C=x_C}$ and $\ket{\tilde{x}_T=x_C \oplus x_T}$, where $\oplus$ is addition modulo 2, or equivalently the XOR operation.}
\end{figure}

There are other modes of quantum computation that are not universal, involving subsets of universal gate operations, or certain global gate operations with less control over the entire space of quantum states.  These can be useful for specific routines or quantum simulations that may not demand full universality. Although global adiabatic Hamiltonian quantum computing \cite{Farhi:2000} can be made universal in certain cases \cite{Aharanov:2004}, it is often better implemented as non-universal subroutines for specific state preparation. Quantum annealing models \cite{Kadowski:1998, Das:2008} do not appear to be universal, and there is a current debate over the advantage such models can have over classical computation \cite{Ronnow:2014}.  Gates that explicitly include error, or decoherence processes~\cite{lin2013dissipative}, used to model quantum computer systems interacting with an environment via quantum simulation~\cite{Carr:2019}, we consider outside the scope of this discussion.

Given the continuous amplitudes that define their quantum states (Eq. \ref{eqn:state}), quantum computers have characteristics akin to classical analog computers, where errors can accumulate over time and lead to computational instability. It is thus critical that quantum computers exploit the technique of quantum error correction (QEC) \cite{Calderbank:1996, Steane:1999}, or at least have sufficiently small native errors that allow the system to complete the algorithm \cite{Preskill_quantum_2018}. QEC is an extension of classical error correction, where ancilla qubits are added to the system and encoded in certain ways to stabilize a computation through the measurement of a subset of qubits that is fed back to the remaining computational qubits. There are many forms of QEC, but the most remarkable result is the existence of fault-tolerant QEC \cite{Preskill:1997, Steane:2003}, allowing arbitrarily long quantum computations with sub-exponential overhead in the number of required additional qubits and gate operations.  Qubit systems typically have native noise properties that are neither symmetric nor static, so matching QEC methods to specific qubit hardware noise profiles will play a crucial role in the successful deployment of quantum computers.

The general requirements for quantum computer hardware \cite{DiVincenzo:2000} are that the physical qubits (or equivalent quantum information carriers) must support (i) coherent Hamiltonian control with sufficient gate expression and fidelity for the application at hand, and (ii) highly efficient initialization and measurement.  These seemingly conflicting requirements limit the available physical hardware candidates to just a few at this time. Below we describe those platforms that are currently being built into multi-qubit quantum computer systems and are expected to have the largest impact in the next decade.
As we will see below in a definition of levels of the quantum computer stack and a sampling of vertical implementations and applications, the near-term advances in quantum computer science and technology will rely on algorithm designers understanding the intricacies of the quantum hardware, and quantum computer builders understanding the natural structure of algorithms and applications.

\vspace{\baselineskip}
\section{The Quantum Computer Stack}

\begin{figure}[h]
  \includegraphics[width=0.4\textwidth]{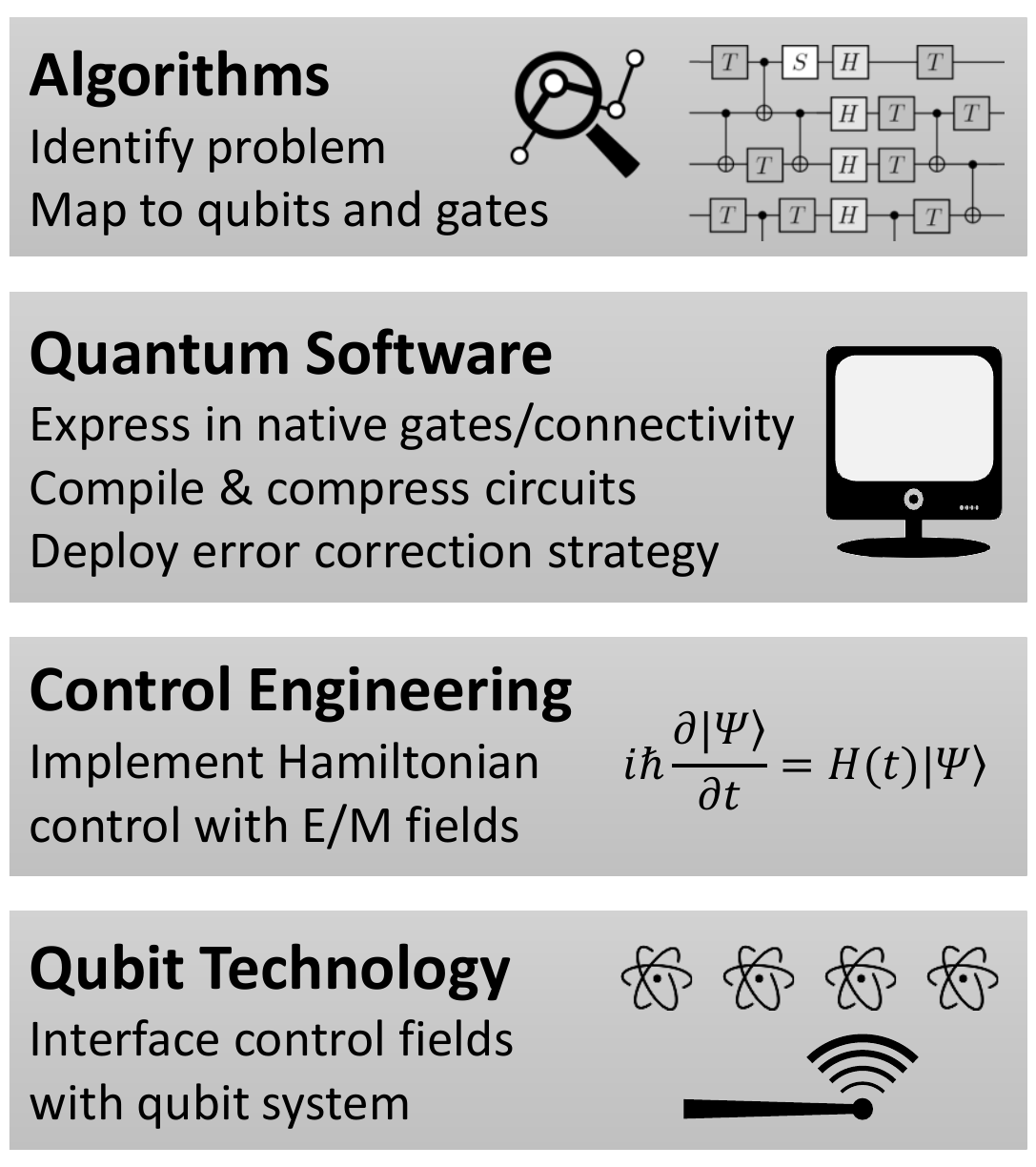}  
  \caption{Advances in all levels of the quantum computer stack, from algorithms and quantum software down to control engineering and qubit technology, will be required to bring quantum computers to fruition. We expect scientific opportunities at every level and at the interfaces between levels.  At the highest levels, quantum computer algorithms are expected to advance many fields of science and technology.  At he middle levels (software stack), the compilation and translation of quantum gates will allow for algorithmic compression to accelerate performance, while error-correction techniques will mitigate quantum computing errors.  At the lowest levels, new ways to control interactions between qubit technologies may lead to better performance.  Future capable qubit technologies will require tight integration with the other layers of the stack to realize their potential. \label{fig:stack}}

\end{figure}

Computer architectures are often defined in terms of their various levels of abstraction or ``stack," from the user interface and compiler down to the low-level gate operations on the physical hardware itself. The quantum computer stack can be defined similarly, as depicted in Fig. \ref{fig:stack}.  However, the various levels of the quantum computer stack (especially the qubits themselves) are not yet cheap and commoditized like classical computer technology.  So it is critical that quantum computers be designed and operated with the entire stack in mind, with a vertical approach of co-designing quantum computer applications to their specific hardware and all levels in between for maximum efficiency.  Indeed, early quantum computer system development may parallel current classical application-specific integrated circuits (ASICs) used for specific and intensive computations such as molecular structure or machine learning.

\sidecaptionvpos{figure}{c}

In this section, we list the levels of the quantum computer stack and point to various approaches at each level.  
The key to co-designing quantum computers is to acknowledge the great opportunities at the interfaces between different levels of the stack, which requires a high level of interdisciplinarity between the physical sciences, engineering, and computer science.
In the next section, we illustrate how various levels of the quantum computer stack will be exploited for several use cases.

\subsection{Quantum Algorithms}

Practical interest in quantum computing arose from the discovery that there are certain computational problems that can be solved more efficiently on a quantum computer than on a classical computer, notably number factoring (Shor’s algorithm~\cite{Shor:1997}) and searching unstructured data (Grover’s algorithm~\cite{Grover:1996}).  Quantum algorithms typically start at a very high level of description, often as a pseudo code. These algorithms are usually distinguished at this level of abstraction by very coarse scaling of resources such as time and number of qubits, as well as success metrics, such as success probabilities or the distance between a solution and the optimal value.  Quantum algorithms can be broadly divided into those with provable success guarantees and those that have no such proofs and must be run with heuristic characterizations of success.

Once a quantum algorithm has been conceptualized with the promise of outperforming a classical algorithm, it is common to consider whether the algorithm can be run on near-term devices or for future architectures that rely on quantum error correction.  A central challenge for the entire field is to determine whether algorithms on current, relatively small quantum processors can outperform classical computers, a goal called quantum advantage.  For fault-tolerant quantum algorithms, on the other hand, there is a larger focus on improving the asymptotic performance of the algorithm in the limit of large numbers qubits and gates.

Shor's factoring~\cite{Shor:1997} and Grover's unstructured search algorithms~\cite{Grover:1996} are examples of ``textbook'' quantum algorithms with provable performance guarantees. These guarantees include a provable quadratic speedup for Grover and a superpolynomial speedup for Shor over known classical algorithms.  A handy guide to known quantum algorithms is the quantum algorithm zoo, \url{https://quantumalgorithmzoo.org/}.  Another important example is the HHL algorithm~\cite{Harrow:2009} which is a primitive for solving systems of linear equations.  Factoring, unstructured search, and HHL are generally thought to be relevant only for larger fault-tolerant quantum computers.

Another class of quantum algorithms are quantum simulations~\cite{Feynman:1982, Lloyd:1996, Carr:2019}, which use a quantum computer to simulate models of a candidate physical system of interest, such as molecules, materials, or quantum field theories whose models are intractable using classical computers. Quantum simulators often determine the physical properties of a system such as energy levels, phase diagrams, or thermalization times, and can explore both static and dynamic behavior. There is a continuum of quantum simulator types, sorted generally by their degree of system control. Fully universal simulators have arbitrary tunability of the interaction graph and may even be fault-tolerant, allowing the scaling to various versions of the same class of problems. Some quantum simulations do not require the full universal programmability of a quantum computer and are thus easier to realize. Such quantum simulators will likely have the most significant impact on society in the short run.  Example simulator algorithms range from molecular structure calculations applied to drug design and delivery or energy-efficient production of fertilizers~\cite{Reiher:2017}, to new types of models of materials for improving batteries or solar cells well beyond what is accessible with classical computers~\cite{saal2013materials}. 

Variational quantum algorithms such as the variational quantum eigensolver (VQE)~\cite{Peruzzo:2014,McClean:2016, Yuan:2019} and the quantum approximate optimization algorithm (QAOA) \cite{Farhi:2014} are recent developments.  Here, the quantum computer produces a complex entangled quantum state representing the answer to some problem, for example, the ground state of a Hamiltonian model of a molecule.  The procedure for generating the quantum state is characterized by a set of classical control parameters that are varied in order to optimize an objective function, such as minimizing the energy of the state.  One particular area of active exploration is the use of VQE or QAOA for tasks in machine learning~\cite{Biamonte:2017} or combinatorial optimization \cite{Yang:2017, Oh:2019}, as discussed below. Variational quantum solvers are a workhorse in near-term quantum hardware, partly because they can be relatively insensitive to systematic gate errors.  However, these algorithms are usually heuristic: one cannot generally prove that they converge. Instead, they must be tested on real quantum hardware to study and validate their performance and compare to the best classical approaches.

Quantum algorithms are typically expressed at a high level with the need to estimate actual resources for implementation.  This often starts with a resource estimate for fault-tolerant error-corrected quantum computers \cite{Terhal:2015}, where the quantum information is encoded into highly entangled states with additional qubits in order to protect the system from noise. Fault-tolerant quantum computing is a method for reliably processing this encoded information even when physical gates and qubit measurements are imperfect. The catch is that quantum error correction has a high overhead cost in the number of required additional qubits and gates. How high a cost depends on both the quality of the hardware and the algorithm under study. A recent estimate is that running Shor’s algorithm to factor a 2048-bit number using gates with a $10^{-3}$ error rate could be achieved with currently known methods using 20 million physical qubits~\cite{Gidney:2019}. As the hardware improves, the overhead cost of fault-tolerant processing will drop significantly; nevertheless, fully fault-tolerant scalable quantum computing is presently a distant goal. When estimating resources for fault-tolerant implementations of quantum algorithms, a discrete set of available quantum gates is assumed, which derive from the details of the particular error-correcting code used.  There are many different techniques for trading off space (qubit number) for time (circuit depth), resembling conventional computer architecture challenges.  It is expected that optimal error correction encoding will depend critically upon specific attributes of the underlying quantum computing architecture, such as qubit coherence, quantum gate fidelity, and qubit connectivity and parallelism.

Estimating resources for quantum algorithms using realistic quantum computing architectures is an important near-term challenge.  Here, the focus is generally on reducing the gate count and quantum circuit depth to avoid errors from qubit decoherence or slow drifts in the qubit control system. Different types of quantum hardware support different gate sets and connectivity, and native operations are often more flexible than fault-tolerant gate sets for certain algorithms. This optimizing of specific algorithms to specific hardware is the highest and most important level of quantum computer co-design.

\subsection{Quantum Software} 

A quantum computer will consist of hardware and software.  Key components of the software stack include compilers, simulators, verifiers, and benchmarking protocols, as well as the operating system.
Compilers --- interpreted to include synthesizers, optimizers, transpilers, and the placement and scheduling of operations --- play an important role in mapping abstract quantum algorithms onto efficient pulse sequences via a series of progressive decompositions from higher to lower levels of abstraction.  The problem of optimal compilation is provably intractable~\cite{janzing2003identity}, suggesting a need for continuous improvement via sustained research and development.  Since optimal synthesis cannot be guaranteed, heuristic approaches to quantum resource optimization (such as gate counts and depth of the quantum circuit) frequently become the only feasible option.  Classical compilers cannot be easily applied in the quantum computing domain, so quantum compilers must generally be developed from scratch.  
Classical simulators are a very important component of the quantum computer stack.  There is a range of approaches, from simulating partial or entire state vector evolution during the computation to full unitary simulation (including by the subgroups of the group of all unitaries), with or without noise.  Simulators are needed to verify quantum computations, model noise, develop and test quantum algorithms, prototype experiments, and establish computational evidence of a possible advantage of the given quantum computation.  Classical simulators generally require exponential resources (otherwise, the need for a quantum computer is obviated) and thus are only useful for simulating small quantum processors with less than 100 qubits, even using high-performance supercomputers.  Simulators used to verify the equivalence of quantum circuits or test output samples of a given implementation of a quantum algorithm can be thought of as verifiers.

Benchmarking protocols are needed to test components as well as entire quantum computer systems.  Quantum algorithm design, resource trade-offs (space vs. gate count vs. depth vs. connectivity vs. fidelity, etc.)~\cite{bishop2017quantum}, hardware/software co-design, efficient architectures, and circuit complexity are examples of important areas of study that directly advance the power of software.

The quantum operating system (QOS) is the core software that manages the critical resources for the quantum computer system. Similar to the OS for classical computers, the QOS will consist of a kernel that manages all hardware resources to ensure that the quantum computer hardware runs without critical errors, a task manager that prioritizes and executes user-provided programs using the hardware resources, and the peripheral manager that handles peripheral resources such as user/cloud and network interfaces. Given the nature of qubit control in near-term devices which requires careful calibration of the qubit properties and controller outputs, the kernel will consist of automated calibration procedures to ensure high fidelity logic gate operation is possible in the qubit system of choice. 
\subsection{Control Engineering}

Advancing the control functions of most quantum computing implementations is largely considered an engineering and economic problem. Current implementations comprise racks of test equipment to drive the qubit gate operations, calibrate the qubit transitions and related control, and calibrate the measurement equipment. While appropriate for the early stages of R\&D laboratory development, this configuration will limit scalability, systems integration, and applicability for fielded applications and mobile platforms, and affordability and attractiveness for future applications.

The quantum gate operations for most qubit technologies require precise synthesis of analog control pulses that implement the gates. These take the form of modulated electromagnetic waves at relevant carrier frequencies, which are typically in the microwave or optical domain. Depending on the architecture of the quantum computer, a very large number of such control channels might be necessary for a given system. While the advances in communication technologies can be leveraged, these have to be adapted for quantum applications. A significant level of flexibility and programmability to generate the required pulses with adequate fidelity must be designed into the control system for quantum computers.

The advancement of controls in quantum computing will ultimately require high-speed and application-specific optimized controls and processing. This situation mirrors the explosive advancement of the telecommunications industry in its implementation of $100$ to $400+$ Gb/s coherent digital optics formats and integrated RF and microwave signal processing for mobile applications.  In the short run, however, the challenge we face is defining the control functions and features relevant for the target qubit applications and deriving the required performance specifications.  This work is necessary before a dedicated, integrated, and scalable implementation, such as ASIC development, becomes viable.  Near-term applications in quantum computing and full system design activities are critical in identifying and defining these needs. There are strong opportunities to engage engineering communities---in academia, national laboratories, and industry---with expertise ranging from computer architecture to chip design to make substantial advances on this front. To foster such efforts, it will be necessary to encourage co-design approaches and to identify common engineering needs and standards.

Generating the types of signals needed can benefit significantly from digital RF techniques that have seen dramatic advances in the last decade. We envision commercially available chipsets based on field-programmable gate arrays (FPGAs) and ASICs that incorporate ``System On a Chip'' (SoC) technology, where processing power is integrated with programmable digital circuits and analog input and output functions. 

Besides the signal generation necessary for implementing the gates, other control needs include passive and active (servo) stability, maintenance of system operation environment (vacuum, temperature, etc.), and managing the start-up, calibration, and operational sequences. Calibration and drift control~\cite{proctor_detecting_2019} are important to both atomic and superconducting systems, though in somewhat different ways.  In fixed-frequency superconducting systems, maintaining fidelities above $99\%$ requires periodic calibration of RF pulse amplitudes; for tunable transmons, low frequency tuning of magnetic flux is required to maintain operation at qubit sweet spots.  For atomic qubits, calibrating local trapping potentials and slow drifts in laser intensities delivered to the qubits is necessary. These calibration procedures, which can be optimized, automated, and built into the operating mode of the quantum computer system with the help of software controls, are candidates for implementation in SoC technologies. The development of optimal operating procedures for drift control and calibration processes will require innovation at a higher level of the stack. For example,  dynamically understanding how control parameter drifts can be tracked and compensated, and when recalibration is needed (if not done on a fixed schedule) will require high-level integration.

Specific performance requirements will help drive progress, by co-design of engineering capabilities and quantum control needs. For example, there is a need for electronic control systems encompassing: (i) analog outputs with faster than 1 Gsamples/s; (ii) synchronized and coherent output with over 100 channels and extensible to above 1000 channels; (iii) outputs switchable among multiple pre-determined states; and (iv) proportional-integrative-derivative (PID) feedback control on each channel with at least kHz bandwidth. Common needs for optical control systems include (i) phase and/or amplitude modulation with a bandwidth of $\approx 100 \operatorname{MHz}$; (ii) over 100 channels and extensibility to above 1000 channels; (iii) precision better than 12 bits (phase or amplitude); and (iv) operating wavelengths to match qubit splittings.

An essential consideration of the control engineering for high-performance quantum computers is noise. The noise in a quantum system has two distinct sources: one is the intrinsic noise in the qubits arising from their coupling to the environment, known as decoherence, and the other is the control errors. Control errors can be either systematic in nature, such as drift or cross-talk, or stochastic, such as thermal and shot noise on the control sources. The key is to design the controller in a way such that the impact of stochastic noise on the qubits is less than the intrinsic noise of the qubits, and the systematic noise is fully characterized and mitigated. Possible mitigation approaches include better hardware design, control loops, and quantum control techniques.

A key element in optimizing control is the characterization of the quantum system.  For small systems, full characterization is possible, but as the system size grows, this becomes impractical. For single and two-qubit gates methods like randomized benchmarking \cite{Emerson_2005,KnillPRA2008,DankertPRA2009} and gate-set tomography \cite{MerkelPRA2013,RBKGST2013} can be used to determine the quality of gates with a tradeoff between information and speed of the benchmark.  For larger systems, system level benchmarks have been developed based on random circuits \cite{bishop2017quantum,Arute:2019}. It is not clear what the best approach for characterizing quantum systems whose scale prevents direct simulation.  Research into characterization of quantum systems is an important component for being able to optimize system level control.     

\subsection{Qubit Technology Platforms}
We view the various quantum computer technology platforms in terms of their ability to be integrated into a multi-qubit system architecture. 
To date, only a few qubit technologies have been assembled and engineered in this way, including superconducting circuits \cite{Havlicek:2019, Arute:2019}, trapped atomic ions \cite{Erhard:2019, Wright:2019, Zhang:2017}, and neutral atoms \cite{Bernien:2017, Levine:2019}.  While there are many other promising qubit technologies, such as spins in silicon, quantum dots, or topological quantum systems, none of these technologies have been developed beyond the component level.  The research and development of new qubit technologies should continue aggressively in materials/fabrication laboratories and facilities.  However, their maturity as good qubits may not be hastened by integrating them with the modular full-stack quantum systems development proposed here, so we do not focus this roadmap on new qubit development.  In any case, once alternative qubit technologies reach maturity, we expect their integration will benefit from the full-stack quantum computer approach considered here.

It is generally believed that fully fault-tolerant qubits will not likely be available soon.  Therefore, specific qubit technologies and their native decoherence and noise mechanisms will play a crucial role in the development of near-term quantum computer systems.
There are several systems-level attributes that arise when considering multi-qubit systems as opposed to single- or dual-qubit systems.  Each of these critical attributes should be optimized and improved in future system generations:
\begin{itemize} 
\item \textbf{Native quantum gate expression.}  Not only must the available physical interactions allow universal control, but high levels of gate expression will be critical to the efficient compilation and compression of algorithms so that the algorithm can be completed before noise and decoherence take hold.  This includes developing overcomplete gate libraries, as well as enabling single instruction, multiple data (SIMD) instructions such as those given by global entangling and multi-qubit control gates \cite{Maslov:2018}. 
\item \textbf{Quantum gate speed.} Faster gates are always desired and may even be necessary for algorithms that require extreme repetition, such as variational optimizers \cite{Farhi:2014, McClean:2016, Yuan:2019} or sampling circuits \cite{Aaronson:2013}. However, faster gates may also degrade their fidelity and crosstalk, and in these cases, the speed to complete the higher level algorithmic solution should take precedence.
\item \textbf{Specific qubit noise and crosstalk properties.} Qubit noise properties should be detailed and constantly monitored, for there are many error mitigation techniques for specific or biased error processes that can improve algorithmic success in the software layer.  Quantum gate crosstalk is usually unavoidable in a large collection of qubits, and apart from passive isolation of gate operations based on better engineering and control, there are software solutions that exploit the coherent nature of such crosstalk and allow for its cancellation by design.
\item \textbf{Qubit connectivity.}  The ability to implement quantum gate operations across the qubit collection is critical to algorithmic success.  While full connectivity is obviously optimal, this may not only lead to higher levels of crosstalk, but ultimately resolving the many available connection channels may significantly decrease the gate and algorithmic speed.  Such a trade-off will depend on details of the algorithm, and a good software layer will optimize this trade-off.  A connection graph that is software-reconfigurable will also be useful.
\item \textbf{High level qubit modularity.} For very large-scale qubit systems, a modular architecture may be necessary \cite{Monroe:2013, Devoret:2013, Monroe:2016}.  Just as for multi-core classical CPU systems, the ability to operate separated groups of qubits with quantum communication channels between the modules will allow the continual scaling of quantum computers.  Modularity necessarily limits the connectivity between qubits, but importantly allows a hierarchy of quantum gate technologies to allow indefinite scaling, in principle.
\end{itemize}

Increasing the number of qubits from hundreds to thousands will be challenging because current systems cannot easily be increased in size via brute force. Instead, a new way of thinking on how to reduce the number of external controls of the system will be needed to achieve a large number of qubits. This could be approached by further integrating control into the core parts of the system or by multiplexing a smaller number of external control signals to a larger number of qubits. In addition, modularizing subsystems to be produced at scale and integrating these into a networked quantum computer may well turn out to be the optimal way to achieve the necessary system size~\cite{Monroe:2016, Loncar:2019}.  Many challenges and possible solutions will only become visible once we start to design and engineer systems of such a size, which will, in turn, be motivated by scientific applications. 

\vspace{\baselineskip}
\section{Quantum Computer Case Studies}

Below we briefly illustrate the use of the quantum computer full stack in several case studies, from optimization problems and programmable simulations of quantum problems to quantum error-correcting codes and ``textbook” algorithms.  These examples are not meant to be exhaustive, but may indicate how any future quantum application might be realized and co-designed.  Mapping these problems onto particular quantum computing modes and specific hardware platforms illustrates the critical translation from cost functions or hard problems to native interactions between qubit systems.  A full stack systems approach to any quantum computation is also expected to inform other scientific applications, even those not yet discovered. In the below use cases, we highlight particular computing modes and qubit technologies that are available now, are expected to scale up significantly in the next 2-10 years, and appear well-matched to the application.

\subsection{I. Gate-Based Quantum Simulation}

The brute-force classical simulation of $n$ interacting qubits requires a solution to $2^n$ complex differential equations, limiting classical computers to simulate arbitrary dynamics on no more than about $50$ qubits. One of the most promising near-term applications of quantum computers is thus the simulation of difficult quantum Hamiltonian models \cite{Feynman:1982} such as frustrated magnetism, superconductivity, and topological dynamics in condensed matter physics or quantum field theories in nuclear and high energy physics. There are many specialized approaches to quantum simulation \cite{Carr:2019} that may have limited tunability but may be easier to implement because of symmetries or natural interactions in the experimental system.  
However, the flexibility of universal gate-based quantum simulators may allow the full power of quantum computers to solve entire classes of models without necessarily specializing in particular cases.

General Hamiltonian simulation evolves an initial state according to the given Hamiltonian model of the system under study. The goal is to minimize the gate count as a function of system size, evolution time, desired precision, and other parameters. There is a wide range of polynomial-time quantum algorithms that solve this problem. They roughly fall into two types: product formulas \cite{Lloyd:1996, Suzuki:1985, Huyghebaert:1990, Berry2007, Childs:2018}, and linear combinations of unitaries \cite{Berry:2015, Low:2017}.  
Apart from theoretical studies aiming to discover new efficient quantum algorithms and improve asymptotic upper bounds, there are also resource count and empirical performance studies \cite{Childs:2018, Babbush:2018, Nam:2019b} as well as approaches that take advantage of the spatial locality of the Hamiltonian to reduce the gate count \cite{Haah:2018, Tran:2019}. 
 
Recent results have provided strong theoretical evidence that sampling from the distribution obtained by evolving some initial state with a Hamiltonian and then measuring cannot be solved by any polynomial time classical algorithm \cite{aaronson:2011}.  This suggests that the simulation of quantum dynamics is a problem well-suited for a solution by a quantum computer.  The major open problem is to understand how robust this exponential speedup is to experimentally relevant noise.

Here we consider universal gate-based approaches to quantum simulation, in its original spirit of attacking a general class of quantum problems \cite{Feynman:1982}.  We assume that a model quantum Hamiltonian is expressed in terms of controllable interactions or gates on a collection of qubits, allowing the simulation of arbitrary quantum processes \cite{Lloyd:1996}.  Such universal quantum simulations can be used to find equilibrium properties of arbitrary Hamiltonians but also allow the more difficult problem of evolving them in time for simulations of quantum dynamics and nonequilibrium processes in physical systems.  Below we consider two gate-based quantum simulation applications: variational simulations of Hamiltonian ground states and the simulation of quantum field theories.

\subsubsection{\textit{(a) Variational Estimation of Ground States}}  
For many problems in quantum chemistry and materials science, it is of great interest to understand the structure of the system electronic ground state or thermal state under a given Hamiltonian. While preparing the ground state of a general Hamiltonian (even if it is spatially local) is believed to be hard even on a quantum computer \cite{kitaev99,oliveira08}, there are algorithms for preparing ground states in special cases \cite{Wecker:2015, Jiang:2018}.  Quantum Approximate Optimization Algorithms (QAOA) \cite{Farhi:2014}, discussed further in the next section, may also be used to heuristically approach Hamiltonian ground states \cite{Ho:2019}.

\begin{figure}[h]
  \includegraphics[width=0.5\textwidth]{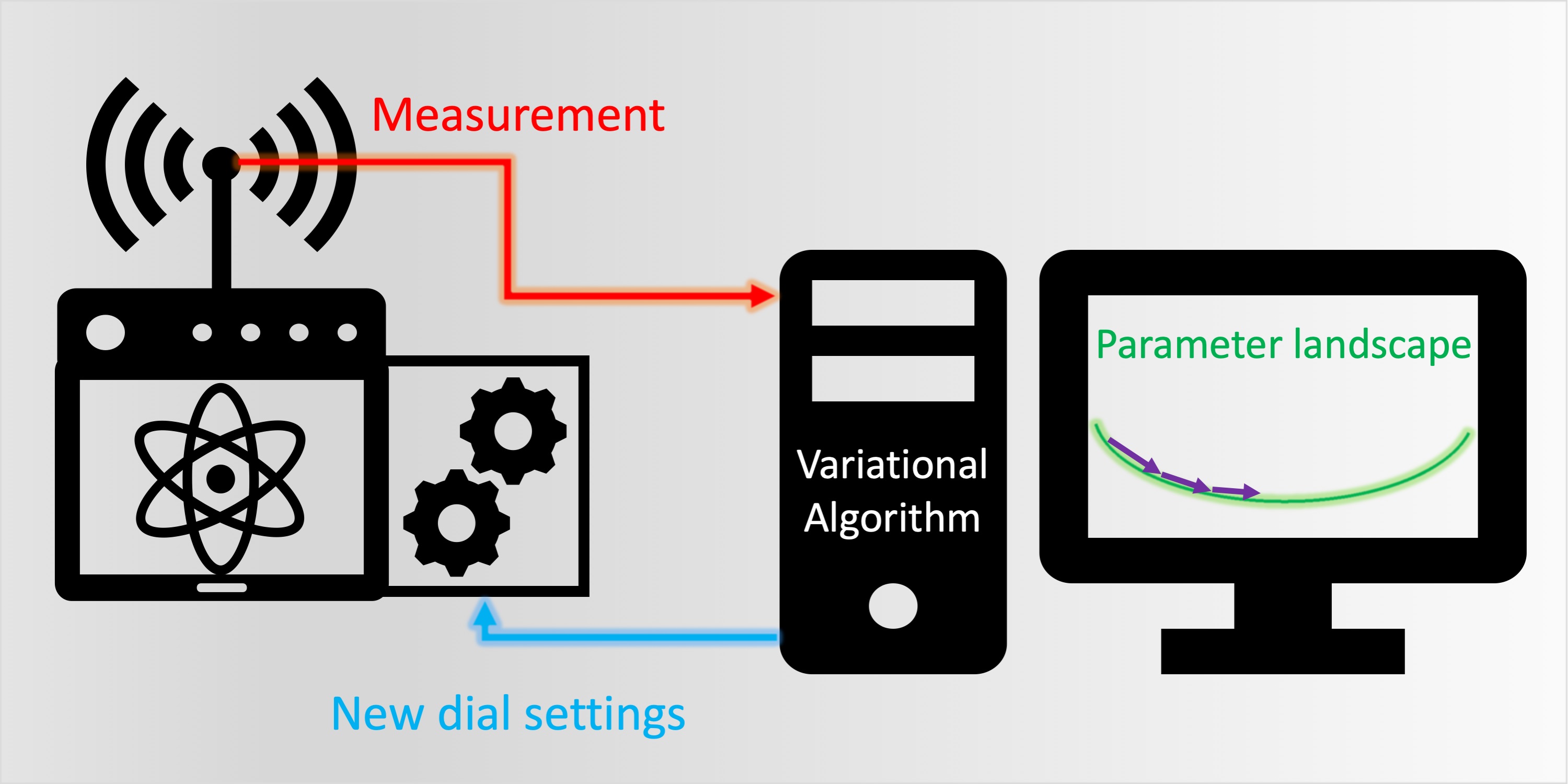}
  \caption{\label{Fig:VQE} Schematic of the measurement and feedback optimization in the variational quantum eigensolver algorithm (from Ref. \cite{Yuan:2019}).}
\end{figure}

Here we concentrate on the implementation of the VQE algorithm \cite{McClean:2016}  of an electronic Hamiltonian using a gate-based approach. The general procedure of VQE involves two steps: The first (preparation) step maps the problem Hamiltonian to a collection of qubits.  In cases such as electronic structure in molecules or materials, this step typically involves a transformation from the native fermionic electron operators to spin or qubit operators through the Jordan-Wigner or Bravyi-Kitaev transformation \cite{Tranter:2018}. The binary occupancy of a fermionic lattice site is replaced by an effective qubit through this substitution.  This transformation results in nonlocal ``string" operators between the effective qubits \cite{McClean:2016, Tranter:2018}, which typically represent the most expensive part of the quantum circuit and are best expressed with highly connected qubit graphs \cite{Nam:2019}. The second (operational) step directly evaluates the Hamiltonian expectation with respect to an initial quantum state parameterized in terms of variables that are classically optimized to minimize the Hamiltonian function, as depicted in Fig. \ref{Fig:VQE}.


\subsubsection{\textit{(b) Simulating Dynamics in Quantum Field Theories}}
An important class of gate-based quantum simulation is the modeling of classically intractable dynamics of quantum field theories (QFTs) at the heart of many phenomena in condensed matter, high-energy, and nuclear physics.
Quantum simulations of QFT dynamics could shed light on many important scientific problems, including phase transitions in the early universe, the response of new exotic materials, explosive astrophysical events and the inner structure of neutron stars, studies of topological features in quantum systems, and high-energy collisions used to search for new fundamental physics.

For scalar quantum field theories, such as those used to describe electron densities in materials or spontaneous symmetry breaking responsible for generating the masses of subatomic particles, the resource estimates for mapping the problem onto qubit registers and the entangling gates have been performed~\cite{Jordan:2011ci,Jordan:2011ne,Klco_2019}. Early estimates for a one-dimensional (1D) lattice scalar field theory shows that some small-scale problems can be mapped onto quantum computers that might become available within the next decade.
Non-Abelian gauge field theories require even more control, although the first SU(2) calculations has been performed on IBM's QExperience~\cite{Klco:2019evd}.  
Resource estimates for qubits and entangling gates are currently being determined for both SU(2) and SU(3) models. The resource requirement estimates are complicated by the need to explicitly enforce Gauss's law.
Quantum chromodynamic (QCD) theory with six flavors of quarks --- each with four Dirac degrees of freedom and three colors --- requires a much larger number of qubits per site. Low-dimensional quantum field theories that share features of the Standard Model are beginning to be explored with available quantum hardware \cite{Martinez:2016,Klco:2018kyo,Lu:2018pjk,Yeter-Aydeniz:2018mix,Klco:2019xro,Davoudi:2019bhy}.


Gate-based universal quantum simulators may be preferable over analog simulation methods for QFT calculations in the longer term, given the need for quantifying uncertainties. However, there is significant value in performing simulations on near-term devices until error-corrected quantum computers become available for simulation. Computations performed on present-day superconducting devices are limited by the qubit coherence time, and the time-dependent behavior of the communication fabric. Measurement errors can be mitigated for modest-sized Hilbert spaces, but this may become challenging at larger scales due to the required classical computing resources.

Given the current status of the hardware capabilities and the resource requirements for QFT methods, co-design is expected to play a key role in closing the gap. 
Starting in the 1980s, co-design played an essential role in developing high performance computing (HPC) capabilities for high-energy physics theories, such as calculating lattice QCD models. An example is the co-design team of Columbia--Brookhaven--IBM that led the design and development of customized computational engines for QCD. In their approach, RAM was integrated onto the floating point unit, with a four-dimensional toroidal communication fabric.  
This effort led to IBM’s Blue-Gene/L,P,Q series of supercomputers.  We anticipate that similar co-design efforts will be required for the development of quantum devices for lattice QFT, e.g., for an efficient evaluation of the sequence of Trotter steps required in time evolution. 

This task requires a close collaboration among a number of experts in a range of areas.  First, particle, nuclear, and condensed matter theorists with expertise in QFT and phenomenology are required to prioritize target observables.  These problems must be mapped onto target hardware, initial states prepared and evolved forward in time, which requires efforts from quantum computer scientists, developers, and engineers with detailed knowledge and hands-on access to the devices.  Circuit optimization will be key in the near term to accomplish the goals of the project. 

Over the course of the next decade, we expect new quantum algorithms to be developed to tackle problems including real-time dynamics, scattering and inelastic processes in low-dimensional gauge theories, such as the 1D Schwinger model and $Z_2$ models, interacting scalar field theories and SU(N) gauge theories, and subsequently extended to higher dimensions. Early implementations on advanced quantum systems are expected to lead to new physical insights in reliable calculations of dynamic SU(2) and QCD processes, including at finite density, into new and exotic  materials, and into the design of quantum memories for quantum devices. 


\subsubsection{\textit{(c) Physical Platform}}

For the above gate-based quantum simulations, we focus on the trapped ion quantum computer architecture \cite{Hempel:2018, Nam:2019} owing to its high gate fidelities, long-range connectivity and flexible gate expression. Other architectures such as neutral atoms \cite{Banuls:2019bmf}), and superconducting qubits \cite{Omalley:2016, Kandala:2017, Colless:2018} can also be considered as potential leading candidates.
The noise in trapped ion experiments appears to scale sublinearly~\cite{Linke:2017}, indicating that the dominant source of error may be due to a coherent source, which can be suppressed through calibration procedures and pulse-shaping techniques. This allows for larger quantum circuit depths and permits an extension of the practical lifespan of computational experiments, compared to physical platforms where the error is dominated by native qubit decoherence. 

More flexible and expressive gate sets allow a given unitary operation to be compiled down to a shorter depth quantum circuit, thus increasing performance. Trapped ion quantum computing allows parallel non-overlapping gates~\cite{figgatt_parallel_2019}, native long-range interactions (all-to-all connectivity that obviates $\swapgate$ operations), small-angle two-qubit rotations, and global gates that act on multiple qubits \cite{lu_global_2019}.  The above features are instrumental in pushing the ability to implement deep algorithms for gate-based quantum simulations and other applications in practice.  

\begin{figure}[htbp]
\begin{align*}
\mbox{\small \Qcircuit @C=0.5em @R=1.4em {
&\qw &\ctrl{1} &\qw &\ctrl{1} &\qw &\qw\\
&\qw &\targ    &\gate{\rzgate(d\theta)} &\targ    &\qw &\qw }}
\hspace{2mm}\raisebox{-4.5mm}{$=$}\hspace{2mm}
\mbox{\small \Qcircuit @C=0.5em @R=0.7em {
&\gate{\rygate(-\pi/2)}  &\multigate{1}{\xxgate(d\theta/2)}     &\gate{\rygate(\pi/2)}   &\qw \\
&\gate{\rygate(-\pi/2)}  &\ghost{\xxgate(d\theta/2)} &\gate{\rygate(\pi/2)}           &\qw  }}
\end{align*}
\caption{In the variational quantum eigensolver algorithm commonly applied to chemistry simulations, small changes in the quantum state induced by controlled rotation involve $\cnotgate$ gates between qubits (left).  However, with the Ising or $\xxgate$ gate native to ion trap quantum computers \cite{Linke:2017}, these small controlled rotations can be expressed as a small-angle Ising gate accompanied by single-qubit rotation operations as written. This expression greatly reduces noise \cite{Nam:2019}.  The rotation operations are $R_x(\theta) = R(\theta,\phi=0)$, $R_y(\theta) = R(\theta,\phi=\pi/2)$ and $R_z(\theta) = R_x(\pi/2)R_y(\theta)R_x(-\pi/2)$.}
\label{Fig:GateReduction}
\end{figure}
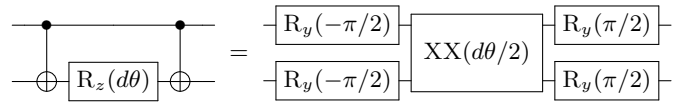

A good example of this reduction in circuit complexity is a key subroutine in a variational quantum algorithm that takes small steps through the Hilbert space, as shown in Fig.~\ref{Fig:GateReduction}.  Here, rotation operations with a small angle $d\theta$ sandwiched by the $\cnotgate$ gates can be replaced by full rotation operations and entangling Ising $\xxgate$ gates evolving with a small angle.  Because the single-qubit rotations are generally much higher quality than the entangling gates in ion traps, translating the small evolution time from the rotation gates to the entangling gates leads to significantly lower errors \cite{Nam:2019, Nam:2019b}. 

Many gate-based quantum simulations have been performed with ion trap quantum computers. These include the simulation of many-body spin dynamics~\cite{lanyon_universal_2011}, real-time dynamics of a lattice gauge theory (the Schwinger model)~\cite{Martinez:2016}, and simulations of Fermi-Hubbard ground states \cite{Linke:2018}. 
To make efficient use of quantum resources, these demonstrations map the original problem to a spin model by exploiting the native long-range interactions, which can be directly and efficiently implemented on an ion trap architecture. 
The second class of gate-based simulations based on the VQE algorithm has been performed with trapped ion quantum computers to calculate the ground state of molecules, from H$_2$ and LiH \cite{Hempel:2018} to H$_2$O \cite{Nam:2019}.  
Related experiments calculated the binding energy of the deuteron (bound proton and neutron) in both superconducting \cite{Dumitrescu:2018} and ion trap systems \cite{Shehab:2019}.

\subsection{II. Combinatorial Optimization with QAOA}

Combinatorial optimization describes a broad class of problems aimed at minimizing a cost function over a combinatorially large set of possible solutions. Examples include problems such as route finding (e.g., the traveling salesman and other graph optimization problems), cost minimization, and portfolio optimization. Optimization applications may ultimately become the most general use case for quantum computers, as they appear in nearly all areas of the natural sciences, engineering, and social sciences.

Here we concentrate on the use of QAOA \cite{Farhi:2014} to solve a particular graph optimization problem. While QAOA can be implemented on a universal gate-based quantum computer, the simple form of its dynamics suggests that QAOA could be implemented on near-term quantum information processors using simpler and more direct techniques than are required for building universal systems. In what follows, we discuss the implementation of the QAOA on the platform of optically-addressed Rydberg atoms. This platform offers different forms of interactions and controls and so is suitable for various classes of geometries for the graph of interactions in the problem Hamiltonian. This point is crucial: even though any NP-complete problem can be encoded in the ground state of a programmable Ising model, and even with nearest-neighbor interactions on a two-dimensional lattice, the extra overhead required to map any given problem onto such an architecture can be large, which can strongly limit the class of problems that can be encoded with a given pattern of interactions. The long-range and non-local interactions available in the Rydberg atom system can greatly expand the range of problems that can be addressed using a device that implements quantum approximate optimization. Identifying optimal embeddings of the problem to be solved into particular architectures of connections is a widely studied problem in quantum software.

\subsubsection{\textit{(a) Maximum Independent Set Problem}}

To illustrate these considerations, we focus on a specific graph optimization problem known as the Maximum Independent Set (MIS), a well known problem involving the search for  the maximum set of vertices in a graph that share no connection \cite{Garey:1979}. Finding the MIS is a canonical optimization problem in the NP-hard complexity class. The most advanced classical algorithms designed to solve such problems exactly often display exponential scaling for vertex sizes above ~100. In particular, there are known instances of graphs with N above ~300 for which classical supercomputers cannot find the MIS. The MIS problem has practical applications in areas such as signal routing in ad-hoc wireless networks such as 5G mobile networks, finance, social network analysis, and machine learning.

The MIS problem can be encoded into the Hamiltonian 
\begin{equation} \label{eqn:MIS}
H_\mathrm{MIS}=-\Delta \sum_{v \in V} n_v + U \sum_{(v,w)\in E} n_v n_w ,     
\end{equation}
where each vertex in the set of all vertices $V$ is represented by a qubit, $n_v = |1\rangle_v\langle1|$, $E$ denotes the edges connecting vertices, and $U$ is the energy cost to excite connected vertices. When $U\gg0$, the ground state of the Hamiltonian encodes the solution to the MIS problem specified by the vertices and edges $V,E$. 


\begin{figure}[h]
  \includegraphics[width=0.5\textwidth]{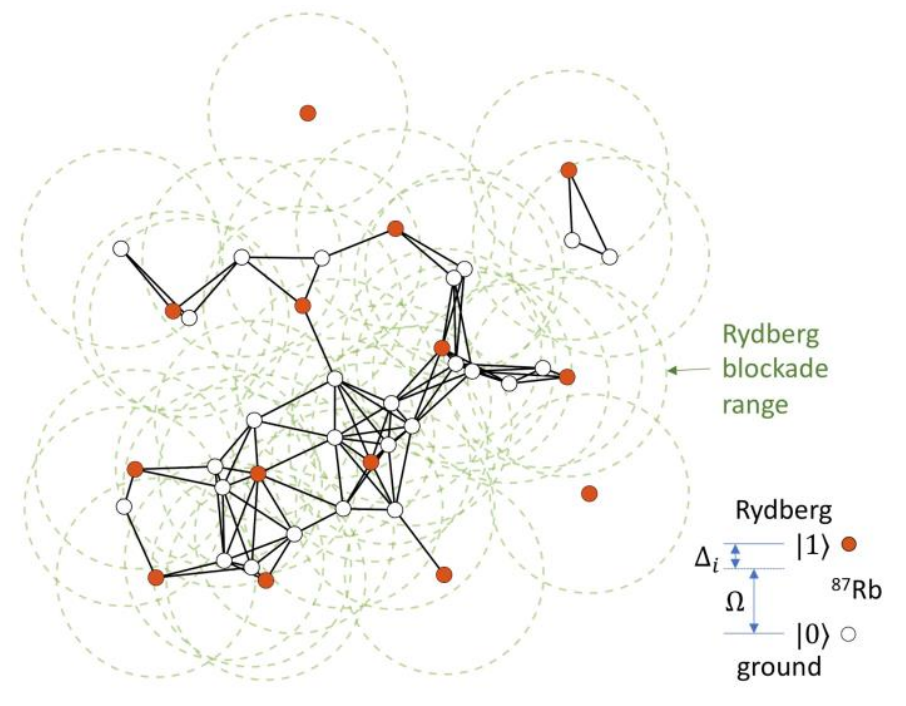}
  \caption{\label{Fig:Graph}Maximum Independent Set (MIS) problem on a random Unit Disk Graph, which is natively represented in an array of atoms with a Rydberg blockade interaction. The MIS of this instance is labeled in red. With variable weights on each vertex, the problem can be generalized to the maximum-weight independent set problem. The legend shows the encoding of qubits in the ground and excited Rydberg state of a neutral atom.  The detuning $\Delta_i$ of a driving field with resonant Rabi frequency $\Omega$ controls the Rydberg blockade radius, which is denoted by the green dashed circles surrounding each atom and is assumed uniform $(\Delta_i = \Delta)$.}
\end{figure}

\subsubsection{\textit{(b) Physical Platform}}

The QAOA for the MIS problems can be efficiently encoded in the Rydberg atom platform. For example, on a subset of graphs, the Unit Disk Graphs (UDGs), the MIS problem can be natively represented with neutral trapped atoms interacting via Rydberg states, without encoding overhead.  As shown in Fig. \ref{Fig:Graph}, every vertex maps directly onto one atom \cite{Pichler2018complexity}, and the edge cost $U$ comes from the Rydberg blockade constraint.  Here, the strong Rydberg--Rydberg interaction impresses a large energy penalty for having two atoms both in excited Rydberg states, represented by the last term of Eq. \ref{eqn:MIS}.

Crucially, the specific MIS problem can be directly encoded in the positions of the atoms (see Fig. \ref{Fig:Graph}), which can be arbitrarily specified in two or three dimensions using reconfigurable optical tweezer arrays based on existing control hardware such as liquid crystal spatial light modulators or digital micromirror devices \cite{Barredo2016,Barredo2018,Kumar2018}. The driving laser pulses that execute the QAOA algorithm can then be global, significantly reducing the control complexity. More general MIS problems can be approached with more sophisticated control and local addressing capabilities. For example, local light-shifts can be used to realize variable weights $\Delta_i$ on each vertex by addressing each atom with a unique driving field frequency and thereby generalizing this situation to the maximum-weight independent set (MWIS) problem \cite{Garey:1979}. Locally programmed interactions using site-resolved excitations to Rydberg states can add additional edges beyond the unit disk graphs, allowing more general instances of MIS problem to be efficiently encoded.

Current demonstrations of optical tweezer arrays for neutral atoms have realized coherent evolution in 51-atom chains \cite{Bernien:2017}, and scaling to larger arrays of 100--1000 atoms is within the reach of current technology. This opens the door to testing for quantum advantage enabled by QAOA for practically relevant problems. Trapped ion systems \cite{Pagano2019} and coherent superconducting circuits \cite{Colless:2018} also represent natural platforms for performing QAOA. Existing superconducting systems support programmable Ising models with local connectivity over 50 superconducting qubits \cite{Arute:2019}, while programmable spin models featuring long-range interactions have been implemented in systems exceeding 50 trapped ion qubits \cite{Zhang2017}. Such systems can also be employed for realizing and testing QAOA on problems of increasing complexity.

Beyond realizing large-scale combinatorial optimization experimentally, there are several open challenges and opportunities in this area. One is to identify how subsets of NP-complete problems such as MIS can be mapped onto other problem classes with near-term devices. For example, how generally applicable would a fast sampling algorithm for the unit-disk MIS problem be? Another is to understand the optimal parameters and Hamiltonian $H_0$ for implementing QAOA for a particular type of problem. A final opportunity is to extend QAOA beyond solving optimization problems, to the rapid preparation of entangled states for other applications \cite{Ho:2019}.

\subsection{III. Quantum Error Correction and Architectures}

Current quantum algorithms are limited by the number of qubits and the quality of gate operations. The physical nature of current qubit technologies and their lack of modularization can limit high level systems engineering and computer architecture approaches to scaling. In this use case, we present a vision for an interdisciplinary architectural goal of the ``virtualization" of qubits, or their abstract representation, through quantum error corrected memories \cite{Terhal:2015}. This activity requires computer architects to design the virtualization strategy, systems engineers to define and design key modules, hardware engineers to improve the classical control and reliability of the system, classical and quantum information theorists to optimize error correction protocols for the platform, and physicists to design new qubits and gate protocols.  The team requires academics to explore the design space, industrial partners to develop equipment and scale promising ideas, and national laboratories to aid in the characterization of materials and testing of devices.  

\subsubsection{\textit{(a) Qubit Virtualization}}

The architecture we envision separates more volatile qubits involved in operations from highly stable qubits stored in memory. This availability of high-quality memory can simplify device design at the cost of requiring more serial operations. We can study the performance of algorithms implemented on this architecture as the ratio of active and memory qubits changes, and estimate the required fidelity of operations and memory for achieving key scientific goals such as simulation of molecules, materials, and high-energy physics. Furthermore, virtualization provides an abstraction to enable computer sciences to use different devices without the need to understand the physical layers.

In the near term, virtualization can be performed using a heterogenous quantum architecture where certain qubits are designed for fast operations, and other qubits are designed for long lifetimes. The cost and benefits of transferring quantum information to and from memory can already be tested and optimized at this stage.  In the long term, we envision that the two qubits could correspond to two quantum error correction codes: for example, a color code for efficient operation \cite{Kubica:2015} and a finite-rate code for efficient data storage \cite{Kovalev:2013}.  The first step in this direction is implementing a fault-tolerant quantum memory with a quantum error correction code.

For universal quantum computing, a broader control space is needed, but the control problem becomes simpler if our goal is to implement a fault-tolerant quantum memory. By optimizing the use of these simple controls, we should be able to study how the performance of the memory improves as it scales up under realistic conditions. 

As the memory is built, we can employ engineering principles to improve the lifetime of the quantum states. A natural strategy is to develop models that track the most likely errors at regular time intervals. Adopting the language of wireless communications, we might refer to this time interval as the channel coherence time and to this process as the channel estimation phase. To optimize for the next time interval, we might develop strategies to efficiently synthesize a quantum error correction code that possesses logical operators that are resilient to those errors identified by the previous channel estimation phase, thereby relaxing full fault-tolerance. Note that the few well-known quantum error correction codes will always be feasible solutions, and our strategy can only do better.

\subsubsection{\textit{(b) Virtualization at the Hardware Level}}
In this use case, we lay out a vertically integrated project to develop virtualized qubits based on physical superconducting qubits combined into fault-tolerant qubits.  We emphasize that both the specific choices of the top architecture and the physical platform are only examples for concreteness. 

Superconducting qubits occupy a large design space with potential tradeoffs among gate speed, coherence time, and readout fidelity \cite{Krantz:2019}.  Recent work has examined transferring quantum information from transmon qubits that can be controlled and processed to cavity qubits with long memory times \cite{Naik:2017}. This design space allows us to test the ideas of modularization and virtualization with devices that can be built immediately.  

Consider, for example, a system of two coupled transmons and two multi-mode cavities.  The cavities can store $N$ qubits of information that can be readily swapped to each transmon, and two-qubit quantum gates can act on the pair of transmons.  Computer scientists can study extended models of the system to examine tradeoffs (with approximately fixed physics parameters) as the number of qubits and cavities increases.  Meanwhile, physicists can continue to develop better methods for transferring information between cavities and transmons \cite{Gao:2019}.  This approach could be combined with bosonic codes that can further extend the  memory times of cavity qubits \cite{Ofek:2016}.

A heterogeneous design for superconducting quantum circuits, combining a variety of qubits and devices, could be more powerful than any single-qubit approach.  One could consider a system that combines transmons, cavity-encoded qubits \cite{Ofek:2016}, ``$0-\pi$" qubits \cite{Gyenis:2019}, and other designs \cite{Krantz:2019},  each with its own role in the system. We envision that this heterogeneous design incorporating noise suppression at the device level will be fed into a quantum error-correcting code arrayed on a two-dimensional lattice, which provides further protection at the software level. For error correction via quantum coding to work effectively,  improvements will be needed in both the underlying physics and the engineering of the control systems.  

Separate from the challenge of memory and more difficult is the implementation of universal fault-tolerant operations. We expect that this is a place where hardware dependent solutions can yield a significant advantage. For superconductors, a promising area of research is to develop fault-tolerant gates in the context of the bosonic codes.  An alternate approach is to design a circuit that implements the light-weight universal gate sets on small block codes. Realizing two universal fault-tolerant qubits is a grand challenge that will stretch our ability to make large and reliable systems.

Architectural requirements for abstract quantum systems place constraints and design requirements on engineered quantum systems.  The construction of systems that achieve these goals requires tight collaboration between computer scientists, engineers, and physicists. 

\subsection{IV. Standard Quantum Algorithms}

Historically, Shor's~\cite{Shor:1997} and Grover's~\cite{Grover:1996} algorithms, which are sometimes referred to as ``textbook algorithms,'' were the first algorithms of practical relevance where a quantum computer offered a dramatic speedup over the best known classical approaches. Here we consider co-design problems and full-stack development opportunities that arise from mapping these textbook algorithms to quantum computing platforms.

At the top of the stack, challenges that both algorithms face include implementing classical oracles with quantum gates. In Grover's algorithm, the oracle can be implemented using Bennett's pebbling game~\cite{bennett_time/space_1989}. However, it is a non-trivial task to find an efficient reversible circuit, since the most efficient implementation on a quantum computer may not follow the structure suggested by a given classical algorithmic description, even when the latter is efficient. There are also intriguing questions in terms of how we can use classical resources to aid in quantum algorithms. For example, in Shor's algorithm we can leverage classical optimizations such as windowed arithmetic \cite{Gidney:2019b}, or trade off quantum circuit complexity with classical post-processing complexity \cite{Ekera:2017,Gidney:2019}.

One of the key challenges in implementing textbook algorithms in physical systems is to optimize these algorithms for a given qubit connectivity and native gate set. While these technology-dependent factors will not affect asymptotic scaling, they could greatly influence whether these algorithms can be implemented on near-term devices. For instance, high connectivity between qubits can provide significant advantages in algorithm implementation~\cite{Linke:2017}. Also, in this vein, work has been done to implement Shor's algorithm with a constrained geometry (1D nearest neighbor \cite{Fowler:2004,Kutin:2006}), but there are many open questions that involve collaborations across the stack.

Developing implementations of Shor's algorithm and Grover's algorithm will provide exciting avenues for improved error correction, detection, and mitigation.
While error correction codes are often designed to correct specific types of errors for particular physical systems, considering error correction for textbook algorithms provides a basis for designing error correction for both application and hardware. Because the output of these textbook algorithms is easily verifiable, they are good testbeds for error mitigation and characterization. For example, simple error correction/mitigation circuits, including randomized compiling~\cite{wallman_noise_2016, campbell_random_2019}, could be implemented in the context of small implementations of Grover’s algorithm and Shor’s algorithm to better understand the challenges in integrating these protocols into more complex ones. While this approach has been used in the quantum annealing community \cite{pearson2019analog}, it could be fruitful to explore in more detail for textbook gate-based algorithms.

Grover’s algorithm seems to break down in the presence of a particular type of error \cite{Regev:2012,Temme:2014} that appears to be unrealistic in actual physical systems. For other algorithms, realistic errors do not appear to be as detrimental \cite{Riste:2015}. Testing textbook algorithms on different architectures with and without error mitigation would provide us with a way to explore the space of errors relative to a specific algorithm, and give insight into which realistic errors are critical. This would inform error mitigation (not necessarily correction) techniques at both the code and hardware level, tailored to specific algorithms. 
This way of viewing error correction calls for a full integration of experts at the hardware, software, and algorithm design levels.

We also expect that the work on these textbook algorithms will lead to improved modularity in the quantum computing stack. In software design, modularity refers to the idea of decomposing a large program into smaller functional units that are independent and whose implementations can be interchanged. Modular design is a scalable technique that allows the development of complex algorithms while focusing on small modules, each containing a specific and well isolated functionality. These modules can exist both at the software level as well as at the hardware level. As an example in classical computing, the increased use of machine learning algorithms and cryptocurrency mining has led to a repurposing of GPUs.

In the design of full implementations of quantum algorithms such as Shor’s and Grover’s, modular design can be applied by using library functions that encapsulate circuits for which an optimized implementation was derived earlier. Examples include quantum Fourier transforms \cite{MikeAndIke, coppersmith:2002approximate, nam:2018approximate}, multiple-control gates \cite{jones:2013low, maslov:2016advantages}, libraries for integer arithmetic and finite fields, and many domain-specific applications such as chemistry, optimization, and finance \cite{McClean:2017,AQUA,Low:2019}. All major quantum computing programming languages are open source, including Google’s Cirq, IBM’s QisKit, Microsoft’s Quantum Development Kit, and Rigetti’s Quil, which facilitates the development and contribution of such libraries. Highly optimized libraries that are adapted to specific target architectures, as well as compilers that can leverage such libraries and further optimize code, are great opportunities for large-scale collaborative efforts between academia, industry, and national labs. 

Like libraries, programming patterns provide opportunities for modularity. Programming patterns capture a recurring quantum circuit design solution that is applicable in a broad range of situations. Typically, a pattern consists of a skeleton circuit with subroutines that can be instantiated independently. Examples of patterns are various forms of Quantum Phase Estimation (QPE) \cite{Cleve:1998, Kimmel:2015, Rudinger:2017, Wiebe:2016}, amplitude amplification \cite{Grover:1996}, period finding \cite{Simon:1997}, hidden subgroup problems \cite{Ettinger:2004}, hidden shift problems \cite{vanDam:2002}, and quantum walks \cite{Venegas:2012}.

Finally, we expect that implementing textbook algorithms will become important benchmarks for the quantum computing stack as a whole and also at the level of individual components.   The need for such benchmarks is evident in the recent proliferation of a variety of benchmarks that test various aspects and components of quantum systems and entire systems, such as quantum volume \cite{Cross:2019}, two-qubit fidelity, cross-entropy \cite{Boixo:2018, Linke:2018, Brydges:2019}, probability of success, reversible computing \cite{maslov2005reversible}, and the active IEEE working group project \cite{IEEE}. It is not likely that any single benchmark will characterize all relevant aspects of a quantum computer system. However, implementing textbook algorithms provides an easy-to-verify test of the full quantum system from hardware to software, as all aspects of the system must work in concert to produce the desired output.

\vspace{\baselineskip}
\section{Outlook and Paths Forward}
The field of quantum computing is now in an era of quantum systems development, where full stack consideration is poised to accelerate progress in building, using, and optimizing quantum computers. The birth and development of quantum computers have taken place in a scientific atmosphere, and we believe that the largest opportunities in quantum computing applications in the coming decade will continue in the realm of scientific discovery.  This stems from the physics of good qubit systems, and the electrical and systems engineering of controlling them to computer science approaches in optimizing the expression of algorithms, finally to be used for nearly all areas of science that will reap the benefits of a new type of computing tool.

\subsection {Community building and engagement}
Quantum computing pulls the rug out from underneath the principles of conventional computing, and its progress will be accelerated by the active engagement of a broad society of users. One mechanism is to underwrite various forms of challenge competitions, which have been successful in stimulating interest and engagement in conventional computing platforms while reaching a large group of coders (e.g., Google Jams, Facebook, and Topcoder competitions). In the quantum computing world, such challenges have already been successfully held, such as the IBM layout and computation challenges, the Microsoft Q\# coding challenges, and the meQuanics quantum circuit minimization challenge. Supporting more challenges like these might help to engage a future quantum workforce around concrete and small scale problems as well as stimulate a cultural shift toward quantum computational thinking. These events can also exploit the growing quantum cloud presence by running routines on real hardware and their virtual simulators.

Universities and national laboratories across the country are setting up quantum information science and technology centers to grow their faculty, unify with their researchers and students, and engage with an increasingly fascinated and enthusiastic general public. These centers, often regional in nature, might play an important role in the future, growing to become large research hubs and advancing state-of-the-art quantum information science. However, they currently play a limited role in consolidating resources, coordinating efforts, and developing collaborative research programs between industry, academia, and national laboratories. All efforts discussed in this roadmap to accelerate the progress in quantum computing should coordinate with these centers of excellence, stimulating the exchange of people and ideas, and growing their potential.

\subsection{Quantum Computing Laboratory User Facilities}

The grand challenges of quantum computing, from finding useful quantum applications to building the machines themselves, are well-motivated by the vast potential scientific and technological opportunities that lie ahead. It is the consensus of the quantum computing community as represented by our workshop that scientific quantum computing user facilities (QCLabs), bringing all of the science, computer science, and engineering of the quantum computer stack together in one place, may best address these challenges and opportunities.  

Each QCLab could have its own type of qubit system, scaling architecture plans, or use-case family, for instance.  But the QCLabs would also provide the capabilities to realize these complex systems and enable collaboration to feature quantum computer co-design up and down the quantum stack, continually iterating on device design, software optimization, and use cases.  These facilities would feature a deep bench of scientists and engineers permanently on site (faculty, staff engineers, etc.), but would also support visitors from all levels: from theorists and algorithmic designers, computer scientists, electrical and computer engineers, to physicists, chemists, and materials scientists.  Each level would be expected to contribute to not just the use of the devices, but also in the continual building of next generations of quantum computers at the QCLab. 

QCLabs are expected to be highly leveraged by industrial efforts in quantum computing. There are many industrial teams now building ever more powerful quantum computing systems, with many of these efforts providing cloud access to their systems. 
We expect the availability of various types of industrial quantum computers to grow rapidly in the coming years.  
These services will have varying characteristics such as qubit platform, qubit number, gate depth, and level of connectivity and expression, that can be exploited by the scientific community.  
However, these cloud services will not likely allow users to dig deep into the stack to potentially optimize the low level control of qubits in order to achieve a particular scientific application, and such systems will not likely be built for scientific research goals having no obvious commercial use.
Moreover, commercial quantum computer cloud providers may not want much flexibility in their system design, as this could degrade high level performance for a widely available and reliable cloud service.  
We envision that the QCLabs would leverage industrial cloud quantum computers in order to assist and benchmark aspects of the QCLab activity itself.  
The industry should be very interested in this interaction, as QCLabs may help find quantum killer applications of the future.

\vspace{\baselineskip}
\section{Acknowledgements and Notes}
We acknowledge useful discussions with Jay Lowell (Boeing, Inc.), David Wineland (University of Oregon), Umesh Vazirani (University of California, Berkeley), and Abhinav Deshpande (University of Maryland).  The participants are thankful to Ms. Samantha Suplee (University of Maryland) for organization of the workshop. This work was supported by the U.S. National Science Foundation and based on a summary of an NSF Convergence Accelerator Workshop on ``Scalable Quantum Computing Laboratories" held on October 21-22, 2019 in Alexandria, VA.  It complements and supports parallel goals of two recent related NSF Convergence Accelerator Workshops: ``Quantum Simulators: Architectures and Opportunities" in the area of quantum simulation \cite{Carr:2019} and ``Quantum Interconnects'' in the area of quantum communication \cite{Loncar:2019}. Any subjective views or opinions that might be expressed in the paper do not necessarily represent the views of the U.S. Department of Commerce, the U.S. Department of Energy, or the U.S.  Government.

\bibliographystyle{apsrev}

\end{document}